\documentclass[twocolumn,twocolappendix,tighten]{aastex63}

\usepackage{amssymb, amsmath}
\usepackage{comment}
\usepackage{hyperref}
\usepackage{amssymb, amsmath}
\def\vb#1{\mbox{\boldmath $#1$}}

\usepackage{here}
\defcitealias{HSC21}{HSC21}
\usepackage{ulem}
\received{????}
\revised{????}
\accepted{????}

\submitjournal{ApJ}

\shorttitle{Saturation level of turbulence}
\shortauthors{Higashi, Susa \& Chiaki}

\begin{document}
\title{Saturation level of turbulence in collapsing gas clouds}

\author[0000-0003-1029-7592]{Sho Higashi}
\affiliation{Department of Physics, Konan University, Okamoto, Kobe, Japan}
\author[0000-0002-3380-5302]{Hajime Susa}
\affiliation{Department of Physics, Konan University, Okamoto, Kobe, Japan}
\author[0000-0001-6246-2866]{Gen Chiaki}
\affiliation{Astronomical Institute, Tohoku University, Aoba, Sendai, Japan}
\affiliation{National Astronomical Observatory of Japan, Mitaka, Tokyo, Japan}

\email{E-mail:shigashi.phy@gmail.com}

\begin{abstract}
We investigate the physical mechanism that decides the saturation level of turbulence in collapsing gas clouds.
We perform a suite of high-resolution numerical simulations following the collapse of turbulent gas clouds with various effective polytropic exponents $\gamma_{\rm eff}$, initial Mach numbers $\mathcal{M}_0$, and initial turbulent seeds.
Equating the energy injection rate by gravitational contraction and the dissipation rate of turbulence, we obtain an analytic expression of the saturation level of turbulence, and compare it with the numerical results.
Consequently, the numerical results are well described by the analytic model, 
given that the turbulent driving scale in collapsing gas clouds is one-third of Jeans length of collapsing core.
These results indicate that the strength of turbulence at the first core formation in the early universe/present-day star-formation process can be estimated solely by $\gamma_{\rm eff}$.
\end{abstract}


\section{Introduction} \label{sec:intro}
Astrophysical fluids are well known to be turbulent and magnetized over a wide range of spatial scales, from planetary systems to galaxies.
Turbulent motion and magnetic fields play important roles for the formation of astronomical objects.
In particular, many earlier studies have shown that both of them crucially contribute to star formation.
In fact, turbulence and magnetic fields affect the fragmentation process of parent gas clouds or accretion disks around protostars, in various environments, from ancient to present-day star formation sites.

The first generation of stars, termed Population III (Pop III) stars, are believed to form in host dark matter minihalos with masses $M\sim 10^6 M_{\odot}$ at redshifts $z \gtrsim 10$ in the $\Lambda$CDM paradigm \citep{Haiman96,Tegmark97,Nishi99, Fuller00,Abel02,Bromm02,Yoshida03}. 
Due to the lack of heavy elements, which are efficient coolants, Pop III forming clouds are more massive and hotter than present-day clouds. As a result, the typical mass of Pop III stars $(\sim 10-1000 M_{\odot})$ should be higher than present-day stars because of the larger fragmentation scale and higher mass accretion rate \citep[e.g.][]{Susa96,Omukai98,Yoshida08}.
However, recent numerical studies for the mass accretion phase have shown that less massive stars $(\sim 1 M_{\odot})$ form through the fragmentation of accretion disks (so-called disk fragmentation) \citep{Clark11Sci,Greif11,Greif2012,Susa13,Susa19,Susa14,Stacy16,Hirano17,Inoue20,Chiaki22}.
Besides, other previous studies have demonstrated that disk fragmentation is promoted by turbulence \citep{Clark11,Riaz18,Wollenberg20} or suppressed by magnetic fields \citep{Machida13,Sharda20,Sadanari21}.
Thus, the strength of turbulence and magnetic fields is crucial for determining the initial mass function (IMF) of Pop III stars.


On the other hand, in present-day galaxies, stars form from rotating molecular cloud cores with typical temperatures of $\sim 10$ K and magnetic fields of a few tens of $\mu \mathrm{G}$, composed of molecular hydrogen ($\mathrm{H_2}$), carbon oxide (CO), ammonia ($\mathrm{NH_3}$), dust grains, etc. 
Depending on the properties of their environments, the cores have various masses from low ($< 10 M_{\odot}$) to high ($> 100 M_{\odot}$) and turbulent velocities from sub- to supersonic.
Accordingly, the masses of present-day stars lie over a wide range with a power-law distribution \citep[e.g.,][]{Salpeter1955,Kroupa01,Chabrier05}.
Such a cloud core distribution could be modeled by magneto-turbulence in parent molecular clouds \citep[e.g.,][]{Padoan02} which is reflected in IMF, while disk fragmentation after the formation of hydrostatic cores may also modify the IMF.
Thus, the turbulence and magnetic fields are also crucial for the star formation process in the present-day universe.

As for the amplification of turbulence, 
previous studies have shown that 
gravitational contraction can take on that task
\citep{Vazques-semadeni98,Robertson2012,Murray15,Birnboim18,Guerrero-Gamboa20,Mandal20,Hennebelle21}.
In our latest study, \cite{HSC21} (hereafter HSC21), 
we numerically followed the evolution of turbulence in collapsing polytropic/barotropic clouds during the collapse phase.
We also obtained an analytic formula that properly describes the growth of turbulence in collapsing clouds.
In these simulations, we also found that the root-mean-square of the turbulent velocity can reach a few times the sound speed and eventually saturates, even when the initial seed field is feeble.
Although we found the saturation level seems to depend on the effective polytropic exponent, we did not step into detailed mechanisms. 

In this paper, we investigate the saturation mechanisms of amplified turbulence by gravitational collapse in detail.
For this purpose, we first perform numerical simulations following the cloud collapse with various initial conditions.
We describe the setup of the collapse simulations in Section \ref{sec:collapse}, and we present our results in Section \ref{sec:results}.
We then theoretically estimate the saturation level of turbulence in Section \ref{sec:estimate} and compare our numerical results and theoretical estimates in Section \ref{sec:comparison}.
Section \ref{sec:discussion} is devoted to discussion.
We summarize the main points in this paper in Section \ref{sec:summary}.

\section{Collapse Simulations}\label{sec:collapse}
Here, we perform a suite of high-resolution three-dimensional collapse simulations by using the N-body/adaptive mesh refinement (AMR) cosmological hydrodynamics code \textsc{enzo} \citep{Enzo, Brummel-Smith19}\footnote{\url{http://enzo-project.org/}}.
This code solves the compressive hydrodynamic equations with the piecewise parabolic method (PPM) in an Eulerian frame using an HLLC Riemann solver.
The basic equations are
\begin{eqnarray}
    \frac{\partial \rho}{\partial t} + \nabla \cdot (\rho \vb{v}) &=& 0,  \label{eq:beq1}\\
\frac{\partial \rho \vb{v}}{\partial t} + \rho(\vb{v}\cdot \nabla)\vb{v} &=& -\nabla P -\rho\nabla\phi,  \label{eq:beq2}\\
\frac{\partial E}{\partial t} + \nabla \cdot \left(E + P \right)\vb{v} &=& -\rho \vb{v} \cdot \nabla \phi -\Lambda + \Gamma. \label{eq:beq3}
\end{eqnarray}

In these equations, $E, ~ \rho, ~ \vb{v}, ~ \phi, ~ \Lambda$, and $\Gamma$ are the total energy density, density, velocity, gravitational potential, radiative/chemical cooling rate, and radiative heating rate, respectively.
The total energy density is given by $E = e + \rho v^2 /2$, where $e$ is the thermal energy density. Remark that we solve Equation (\ref{eq:beq3}) only for a model with primordial chemistry (\S \ref{sec:chemi}) but not for barotropic runs (\S \ref{sec:baro}).
The gravitational potential $\phi$ is obtained from the Poisson's equation below with a Fast Fourier Transform (FFT) technique \citep{Hockney88}
\begin{equation}
    \nabla^2\phi = 4\pi G\rho.
\end{equation}

We generate a gravitationally unstable Bonner-Ebert sphere with a central density $\rho_{\rm peak,0}=4.65\times 10^{-20} ~\mathrm{g ~ cm^{-3}}$, uniform temperature $T_{0}=200 ~ \mathrm{K}$, and cloud radius $r_{\rm c}$ = 1.5 pc as an initial condition.
This is the same initial condition as \citetalias{HSC21}.
The computational domain has a size of 4 $r_{\rm c}$ with periodic boundary conditions.

We start the calculations with a base grid with $256^3$ cells and progressively refine cells as the cloud collapses in order to resolve the Jeans length at least by 128 cells (we refer to this minimum refinement criterion as a `Jeans parameter').
Except for the model with $\gamma_{\rm eff}=1.3$ (explained later), we terminate simulations when the maximum cloud density reaches $10^{-4} ~ \mathrm{g ~ cm^{-3}}$.
By the end of the simulations, the maximum refinement level reaches 27 for $\gamma_{\rm eff} = 1.0$ and 20 for $\gamma_{\rm eff} = 1.25$. The corresponding minimum spatial resolutions are $3.0\times 10^{-5}$ AU and 0.038 AU, respectively.
The maximum levels in the other runs are in the range of 20-27.
We use the \textsc{yt} toolkit \citep{yt}\footnote{\url{https://yt-project.org/}} to analyze the simulation data.

\begin{table*}[htbp]
 \caption{Properties of simulations} 
 \hspace{-1cm}
   \begin{tabular}{lcccc*{5}c}
     \hline \hline
     Name & Mach number & 
     Initial rotation & $\gamma_{\rm eff}$ & turbulence seed & &&& &Remark\\
     \hline
     LowA  & 0.1 & No & All & \textbf{A} & &&& & fiducial\\
     LowB  & 0.1 & No & All except 1.25 and 1.3 & \textbf{B} & &&& & \\
     LowC  & 0.1 & No & All except 1.25 and 1.3 & \textbf{C} & &&& & \\
     LowA w/c  & 0.1 & No & With chemistry & \textbf{A} & &&& &\\
     Middle  & 0.5 & No & 1.09 & \textbf{A} & &&& &\\ 
     Middle w/r  & 0.5 & Yes & 1.09 & \textbf{A} & &&& &\\ 
     High & 1.0  & No & 1.09 & \textbf{A} & &&& &\\ 
     \hline\hline
    \multicolumn{10}{l}{
     ``Low'', ``Middle'', and ``High'' correspond to the strength of initial Mach numbers. `w/c' and `w/r' represent the
     }\\
     \multicolumn{10}{l}{model  with chemical reactions and initial rotation, respectively. \textbf{Note}: we run simulations for $\gamma_{\rm eff}=1.25$
     }\\
     \multicolumn{10}{l}{
     and $1.3$ only with \textbf{seed A} because of high computational cost.
     }
     \\
   \end{tabular}\\
  \label{tab:init2}
\end{table*}

\subsection{Barotropic EoS}\label{sec:baro}
We calculate the temperature evolution of the collapsing gas by using a simplified polytropic model as  \citetalias{HSC21}.
The relation between the pressure $P$ and the density $\rho$ of gas is expressed as
\begin{equation}
    P \propto \rho^{\gamma_{\rm eff}}
\end{equation}
by using an effective polytropic exponent $\gamma_{\rm eff}$.
In this model, we use this relation instead of solving Equation (\ref{eq:beq3}).
We perform simulations for eight $\gamma_{\rm eff}$: 1.0, 1.05, 1.09, 1.1, 1.15, 1.2, 1.25 and 1.3.
The parameters $\gamma_{\rm eff} =$1.0, 1.09 and 1.2 were also employed in \citetalias{HSC21}, which enables us to compare the results directly.
Note that we terminate the simulation when the mean density within the core reaches $10^{-10} ~ \mathrm{g ~ cm^{-3}}$ only for $\gamma_{\rm eff}=1.3$ because of high computational cost.
Here, the models for $\gamma_{\rm eff}=1.09$ and $1.1$ mimic the primordial gas in minihalos \citep{Omukai98}.
In these models, we set the gaseous mean molecular weight $\mu=1.22$.

\subsection{The model with primordial chemistry}\label{sec:chemi}
Although not explicitly calculated in the barotropic EoS, solving the entropy production/reduction by hydrodynamic shock or radiative cooling, may change the amplification/saturation level of turbulence through the baroclinic term of the vorticity equation \citep[e.g.,][]{Wise07}.
In order to investigate the effects of baroclinic term on the evolution of turbulent velocities, we explicitly take into account non-equilibrium chemistry and radiative cooling of primordial gas.
For this, we solve 49 chemical reactions of 15 primordial species ($\mathrm{e,~ H,~ H^{+},~ H^{-},~ H_2,~ H_2^{+},~ He,~ He^{+},~ He^{2+},~ HeH^{+},~ D,}\\
\mathrm{D^{+},~ D^{-},~ HD}$, and $\mathrm{HD^{+}}$) by using \textsc{grackle} \citep{grackle}\footnote{\url{https://grackle.readthedocs.io/}} modified/extended in \cite{Chiaki19} (see \citealt{Chiaki19} in detail).
In this model, $\mu$ is directly calculated from the abundance of the chemical species.
In the high-density region with a number density above $10^{17} ~ \mathrm{cm^{-3}}$,
the time step of non-equilibrium chemical solver becomes prohibitively smaller than the dynamical time, resulting in a huge computational cost.
Hence, in this region, we calculate the abundances of hydrogen and helium by equilibrium chemical solver using the Saha-Boltzman equation.
As the initial condition, we assume the mass fractions of $\mathrm{H_2, ~ H^{+}, ~ H}$, and $\mathrm{He}$ as $x_{\rm H_2}=7.60\times 10^{-4}$, $x_{\rm H^{+}}=7.60 \times 10^{-8}$, $x_{\rm H}=0.759$, and $x_{\rm He}=0.240$, respectively.

\subsection{Initial turbulent flow}\label{sec:realM}
It is known that collapsing gas cores have some degree of turbulent velocities at the onset of run-away collapse (see the reviews of \citealt{Greif15} for primordial clouds and \citealt{McKee07} for present-day clouds).

Additionally, the power spectrum of the turbulent velocity follows the Larson's law ($P(k) \equiv \langle |\vb{v}_{\rm k}|^2\rangle \propto k^{-4}$)
in both present-day \citep{Solomon87} and primordial gas clouds \citep[][]{Prieto11}, where $\vb{v}_{\rm k}$ and $k$ are the velocity in the wavenumber space and a wavenumber, respectively.
$\langle \cdots \rangle$ denotes the average of a quantity.
In order to mimic the turbulent flow of the initial cloud, we give the initial velocity field following the Larson's law, 
by varying the initial root-mean-square Mach number.

Thus, the 1D kinetic energy spectrum $E(k) \propto k^2 P(k)$ is proportional to $k^{-2}$, which has the same power-law index of Burgers (compressible) turbulence \citep{BURGERS1948}
\footnote{Strictly speaking, this initial velocity field is not self-consistent ``real'' turbulence since there is no correlation between density and velocity. See section \ref{sec:limit}.}.

This turbulent velocity field is composed of the natural mixture of solenoidal (divergence-free) and compressive (curl-free) modes with a 2:1 ratio.
It is the reasonable initial condition, given that transverse modes account for 2/3 of the spatial directions in a 3D system, and longitudinal modes account for the remaining 1/3.
Earlier studies showed that star-forming cores have a few percent of rotation energy against gravitational potential \citep[e.g.,][]{Goodman93,Yoshida06}.
Therefore, 
we also perform the simulation for the more realistic case, where the rotational energy is 5\% of the gravitational potential, to investigate the effects of initial rotation on the temporal evolution of the turbulent velocity.

To investigate 
the effect of the initial random seed of turbulence,
we perform simulations with three different initial turbulent seeds (tagged ``A'', ``B'' and ``C'') for $\gamma_{\rm eff}=$ 1.0, 1.05, 1.1, 1.15 and 1.2.
We consider a single seed (A) for $\gamma_{\rm eff}=$1.25 and 1.3 because of the high computational cost.
The properties of each simulation are summarized in Table \ref{tab:init2}.

In our analysis of the evolution of turbulence (Section \ref{sec:results}), we calculate the cell-volume weighted root-mean-square turbulent velocity within a sphere with a radius of half the Jeans length $L_{\rm J}$ (hereafter called the ``Jeans volume''), as well as \citetalias{HSC21}.
To define a cloud core, 
it is necessary to evaluate the Jeans length because the length scale of a collapsing gas core is typically the Jeans length.
However, the Jeans length depends on the mean density and sound speed within the core of a certain radius.
Thus, we need to assess the core radius accurately at first. 
We calculate the core radius iteratively as follows

\begin{enumerate}
    \item First, we cut out the overdense region of $\rho \geq \rho_{\rm th}$, where $\rho_{\rm th}=\rho_{\rm peak}/16$ in this paper.
    $\rho_{\rm peak}$ is the maximum density in the computational domain.
    
    \item Calculate the center of mass $\vb{r}_{\rm c}$ in the region and `tentative' Jeans length $\left( = \sqrt{\pi c_{\rm s,t}^2/G\rho_{\rm mean,t}}\right)$ using cell volume-weighted average density $\rho_{\rm mean, t}$, and sound speed $c_{\rm s,t}$. And then, we assume this Jeans length as tentative radius $r_0$.\label{it:temp_r0}
    
    \item Considering a sphere with a radius of $r_0$ from $\vb{r}_{\rm c}$, calculate tentative Jeans length and radius $r_1$ as well as \ref{it:temp_r0}. \label{it:temp_r1}
    
    \item If $|r_1 - r_0|$ is smaller than the smallest cell width, we terminate this calculation. If not, we substitute $r_1$ to $r_0$ and go back to \ref{it:temp_r1} to repeat until the above condition is fulfilled.
    
    \item We obtain $r_1$ as the core radius $r_1(=L_{\rm J}/2)$ after convergence.
\end{enumerate}

Within the Jeans volume $V_{\rm J}$, the turbulent velocity is defined as follows:
\begin{equation}
    v_{\rm turb}^2\equiv \sum_{\leq L_{\rm J}/2 } \frac{V_i}{V_{\rm J}}(\vb{v}_i-\vb{v}_{{\rm rad},i})^2, \label{eq:vturb}
\end{equation}
where $V_i, ~ \vb{v}_i, ~ \vb{v}_{\rm{rad},i}$ are the cell volume, the velocity, and smoothed radial velocity of the $i$th cell, respectively.
To calculate the turbulent velocity, we estimate the smoothed background radial velocity through the following procedure:
\begin{enumerate}
    \item Calculate the radial velocity profile with $N_{\rm rad}$ radial bins.
    $N_{\rm rad}$ should be smaller than the Jeans parameter so that we can remove the radial velocity fluctuations at the scale of the cell size. Here, we set $N_{\rm rad} = 16$.
    \item Linearly interpolate this profile to estimate the smoothed radial velocity $\vb{v}_{{\rm rad}, i}$ at the position of each cell.
\end{enumerate}

\begin{figure}[htbp]
 \centering
  \plotone{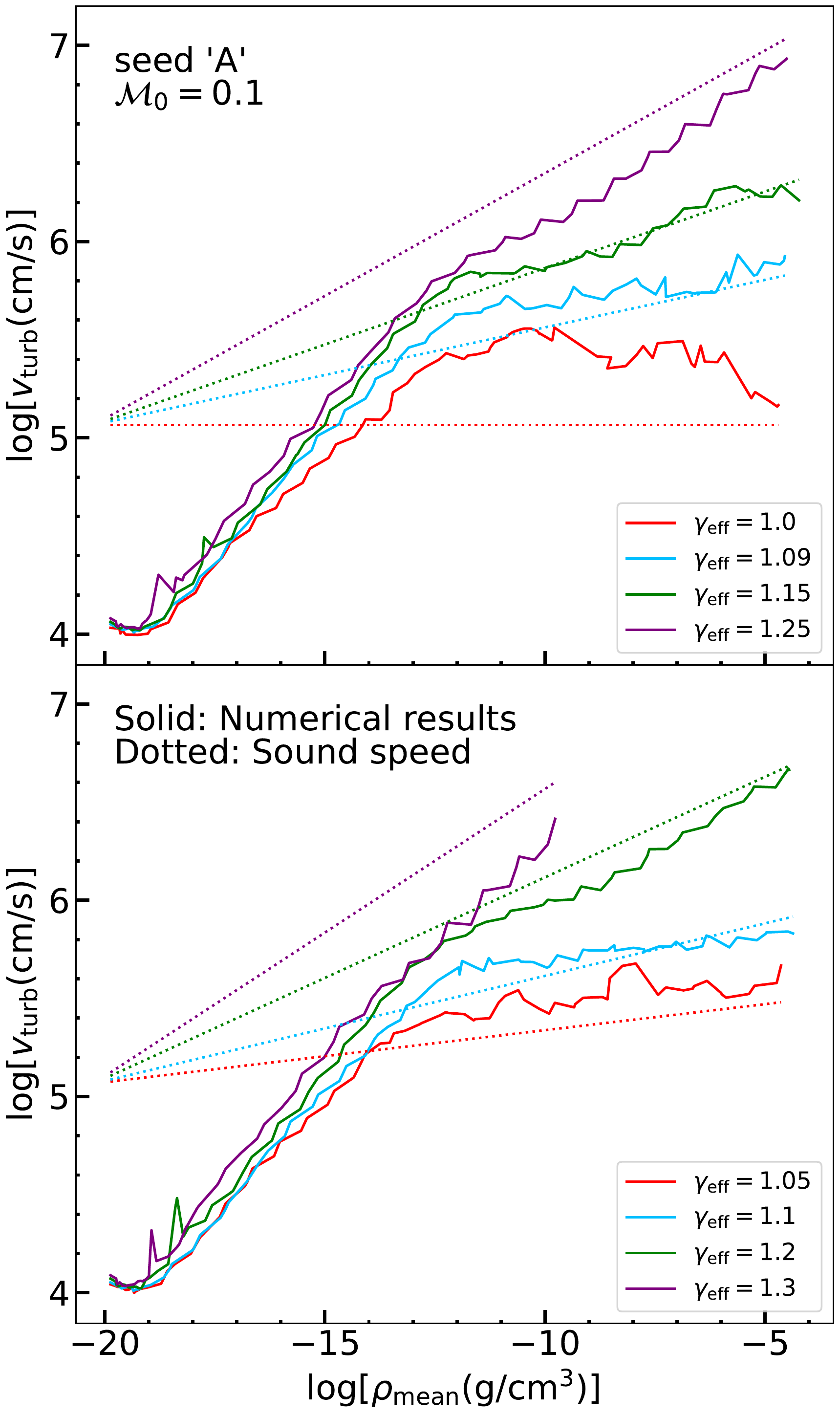}
  \caption{Evolution of the turbulent velocity $v_{\rm turb}$ as a function of mean density $\rho_{\rm mean}$ in the Jeans volume. The solid curves with different colors denote the numerical results for different $\gamma_{\rm eff}$. The dotted lines 
  denote the mean sound speed in the Jeans volume.}
    \label{fig:velocity} 
\end{figure}

\section{Results}\label{sec:results}
In this section, we initially present the overall turbulence evolution in the models with the barotropic EoS focusing on the effects of the polytropic index, initial turbulent seed, and initial turbulent velocity in Section \ref{sec:result_baro}.
We then compare the results with/without the baroclinic term and show its effects on the evolution of turbulence in Section \ref{sec:comparison_chemi}.

\subsection{Overall turbulence evolution in barotropic EoS}\label{sec:result_baro}
Figure \ref{fig:velocity} shows the evolution of the turbulent velocity as a function of the mean density $\rho_{\rm mean}$ within the Jeans volume of each snapshot in the fiducial model (\texttt{LowA}).
Note that we divide the results into two panels for visibility.
As shown in \citetalias{HSC21}, the turbulent velocities (solid curves) increase at the rate that depends on $\gamma_{\rm eff}$ initially. 
Eventually they saturate at $\rho_{\rm mean}\simeq 10^{ -13}~ \mathrm{g ~ cm^{-3}}$ and then increase with the growth rate similar to the sound speed (dashed lines). 
Additionally, the ratio of the saturation velocity to the sound speed depends on $\gamma_{\rm eff}$ and decreases as $\gamma_{\rm eff}$ increases.
The physical mechanism that determines these saturation levels is explained in sections \ref{sec:estimate} and \ref{sec:comparison} in detail. 

\begin{figure}[htbp]
 \centering
  \plotone{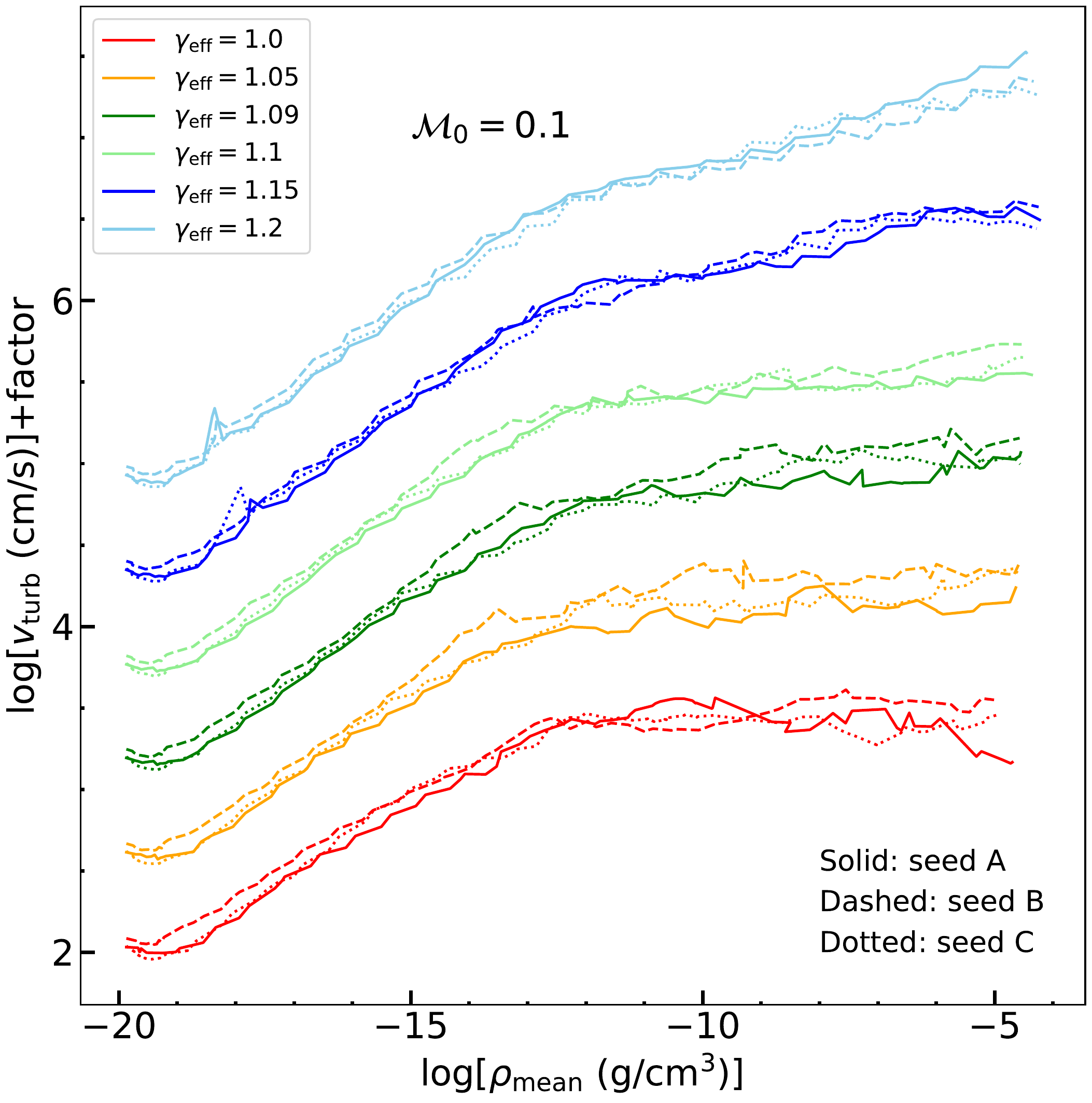}
  \caption{
  Same as Figure \ref{fig:velocity} but for the different initial seeds of turbulence (LowA, LowB, and LowC).
  Difference in line style corresponds to the difference of initial seeds and difference in colors corresponds to the difference of $\gamma_{\rm eff}$.
  For ease of viewing, we multiply the results by factors for each $\gamma_{\rm eff}$ model.
  }
    \label{fig:seed} 
\end{figure}

\begin{figure}[htbp]
 \centering
    \plotone{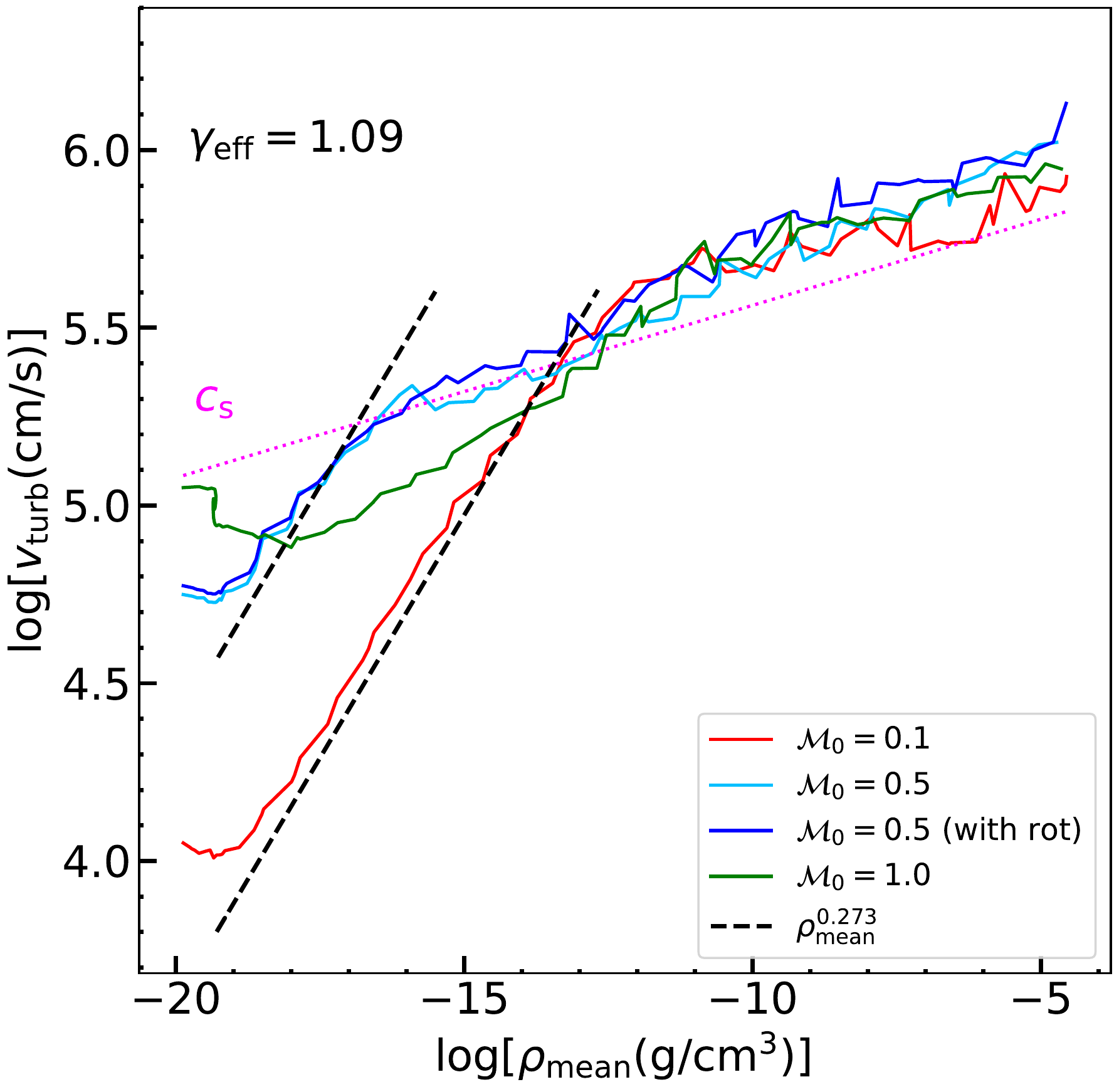}
  \caption{
  Same as Figure \ref{fig:velocity} but for the different $\mathcal{M}_0$ and rotation energies (LowA, Middle, Middle w/r, and High).
  Solid curves are numerical results for each model.
  The magenta dotted line denotes the averaged sound speed in the Jeans volume for $\gamma_{\rm eff}=1.09$. The black dashed line is obtained by an analytic estimate in \citetalias[][Equation (14)]{HSC21}.
  }
    \label{fig:mach} 
\end{figure}

Next, we plot the evolution of turbulent velocities with each initial turbulent seed in Figure \ref{fig:seed}.
For ease of viewing, we multiply the results by factors for each $\gamma_{\rm eff}$.
We can see the growth rates of turbulent velocities before the saturation $\left( \rho_{\rm mean} \lesssim 10^{-13} ~ \mathrm{g ~ cm^{-3}}\right)$ do not depend on the initial turbulent seed.
We can also see that the final turbulent velocities are almost converged in all $\gamma_{\rm eff}$ and the difference in the initial seeds is a factor of two at most.
The difference in Mach numbers among different seeds and its details are explained in Section \ref{sec:comparison}.

Figure \ref{fig:mach} shows the evolution of turbulent velocities for each model with different initial turbulent velocities.
We also plot the results of the model with rotation (Middle w/r).
In all the models, the turbulent velocities saturate at $\rho_{\rm mean}\gtrsim 10^{-13} ~ \mathrm{g ~ cm^{-3}}$.
The saturation levels in these models almost converge and the difference among different initial strengths of turbulence or the effect of rotation are about 20\% at most.
This implies the saturation level of turbulent velocity does not depend on the initial strength of turbulence or the presence of cloud rotation. 

We find that the rotational velocity is subdominant, compared to the turbulent component in our particular models in the entire course of collapse. Thus the resultant Mach number is not affected much by the initial rotation.
However, we have to keep in mind that we must explicitly consider the rotation velocity if it is the dominant component.
In such cases the definition of turbulent velocity (Equation \ref{eq:vturb}) should be modified as Equation (18) in \cite{Chiaki22}.

\subsection{Comparison with/without baroclinic term}\label{sec:comparison_chemi}
\begin{figure}[htbp]
 \centering
  \plotone{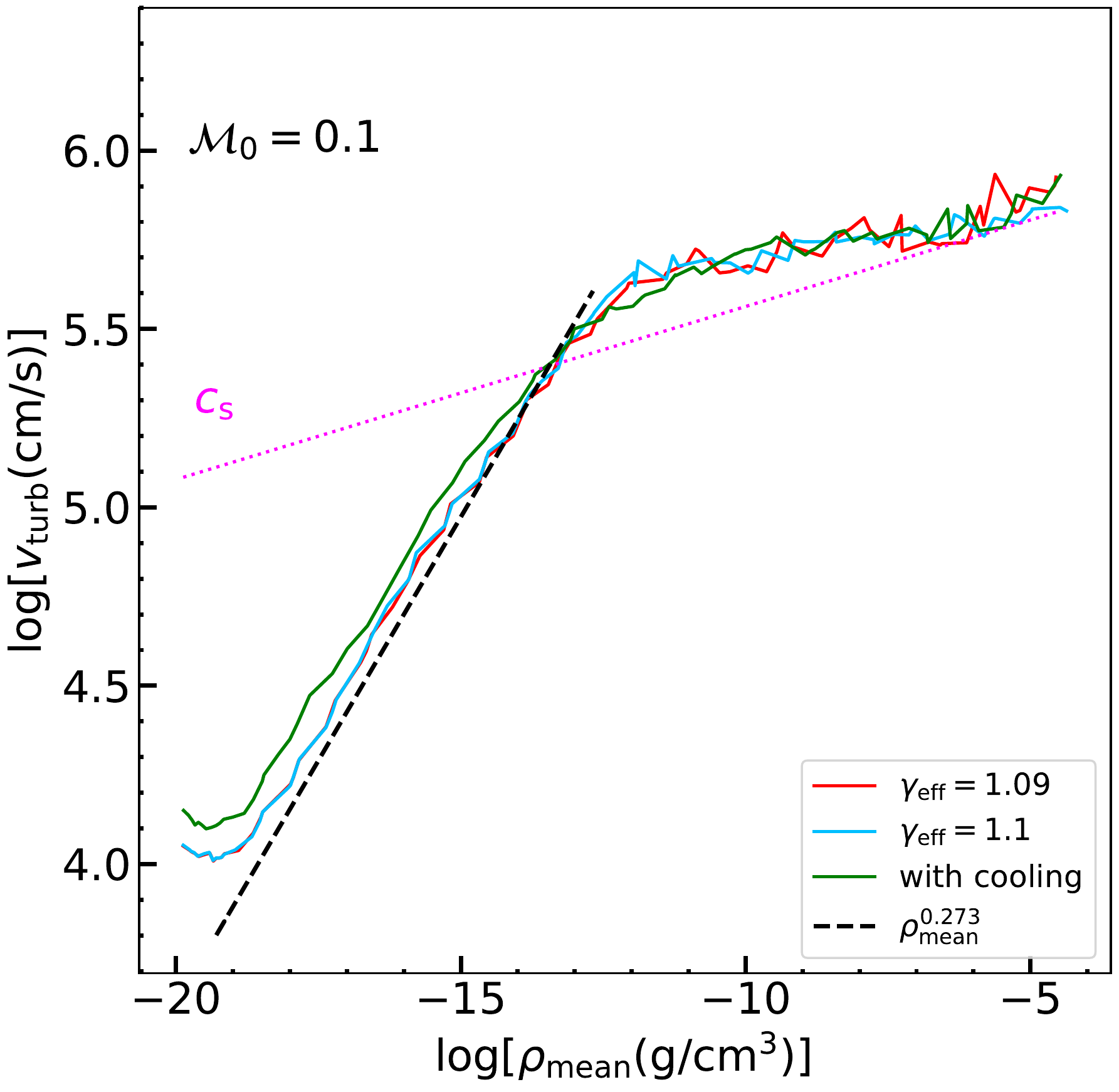}
  \caption{
  Same as Figure \ref{fig:mach} but for barotropic models and the model with primordial cooling.
  The solid curves denote the numerical results for $\gamma_{\rm eff}=1.09$ (red), $\gamma_{\rm eff}=1.1$ (cyan), and the model with radiative cooling (green). 
  The black dashed line is obtained by an analytic estimate in \citetalias[][Equation (14)]{HSC21}.
  }
    \label{fig:comparison_v} 
\end{figure}

As mentioned in Section \ref{sec:baro}, $\gamma_{\rm eff}=1.09$ and $1.1$ well reproduce the temperature evolution of collapsing primordial gas clouds. 
Here we compare the temporal evolution of the turbulent velocities and the saturation levels between the models with the barotropic EoS and the model explicitly treating radiative cooling as shown in Figure \ref{fig:comparison_v}.

As well as the results of barotropic EoS runs (Figure 4) in \citetalias{HSC21}, we can see that all of the results for LowA with $\gamma_{\rm eff}=1.09$, $1.1$, and w/c are in good agreement with the analytic estimate (dashed line) at $\rho_{\rm mean} \lesssim 10^{-13} \mathrm{g ~ cm^{-3}}$.
We can also see that the turbulent velocities in all models saturate at $\rho_{\rm mean} \gtrsim  10^{-13} ~ \mathrm{g ~ cm^{-3}}$ and converge.
This shows that the growth and saturation of turbulence in the gravitational collapse is not affected greatly by the baroclinic term induced by shock waves and chemical reactions/radiative cooling but depends only on $\gamma_{\rm eff}$.

\section{Analytic estimate of turbulence saturation}\label{sec:estimate}

We find the growth and saturation of turbulence depend solely on $\gamma_{\rm eff}$ in Section \ref{sec:results}. 
In this section, extending the analytic model proposed by \cite{Robertson2012}, we analytically estimate the maximum saturation level of turbulence induced by gravitational collapse from the balance between the amplification and the decay due to dissipation of turbulence.

In the ``adiabatic heating'' model in \cite{Robertson2012}, the temporal evolution of rms turbulent velocity on the uniformly contracting background is expressed as 
\begin{equation}
    \frac{dv}{dt}=-Hv -\eta \frac{v^2}{a\ell} \label{eq:RG1}.
\end{equation}
Here, $a$, $H(\equiv \dot{a}/a)$, $\eta$, and $\ell$ are the scale factor, ``Hubble parameter'', dissipation coefficient of turbulence, and the typical driving scale of turbulence in comoving coordinates, respectively.
In the uniformly changing background, the mean density field $\rho$ and $a$ are given by the relation $\rho\propto a^{-3}$.

The first-term on the right-hand side in Equation (\ref{eq:RG1}) denotes the amplification of turbulence due to gravitational collapse, while the second term represents the decay term due to dissipation, respectively.
We can rewrite this equation by using the turbulent eddy turnover frequency $\omega = \frac{v}{a\ell}$ as
\begin{equation}
    \frac{dv}{da} = - \left( 1 + \eta \frac{\omega}{H} \right) \frac{v}{a}. \label{eq:RG2}
\end{equation}

\cite{Robertson2012} also derived the relation between $\omega$ and $H$, from Equations (\ref{eq:RG1}) and (\ref{eq:RG2}),
\begin{equation}
    \frac{d \log(\omega/H)}{d\log(1/a)} =  \left( 2+\eta\frac{\omega}{H}\right)-\frac{d\log H}{d\log(1/a)}. \label{eq:RG_asymptotic}
\end{equation}
Here $\ell$ is assumed to be a constant.
If the collapse time $t_{\rm collapse} \sim |H|^{-1}$ is proportional to the free-fall time ($\propto 1/\sqrt{\rho}$); the second term in Equation (\ref{eq:RG_asymptotic}) becomes
\begin{equation}
    \frac{d \log(H)}{d\log(1/a)} =  \frac{3}{2}. \label{eq:dHda}
\end{equation}
The asymptotic relation is approached as $d\log (\omega/H)/d\log(1/a)~ \rightarrow ~ 0$.
We have 
\begin{equation}
    2+\eta\frac{\omega}{H} -\frac{3}{2} \rightarrow 0 \quad (\mathrm{for} ~ \log(1/a)\rightarrow \infty).
\end{equation}

We apply this model to the collapsing gas core in a self-similar fashion.
Assuming the driving scale in Equations (\ref{eq:RG1}) is a factor $f$ times Jeans length $L_{\rm J}$, and re-deriving the relation corresponding to Equation (\ref{eq:RG_asymptotic}) and (\ref{eq:dHda}), we obtain 
\begin{equation}
    \frac{d \log(\omega/H)}{d\log(1/a)} = \frac{5-3\gamma_{\rm eff}}{2}+\eta \frac{\omega}{H}  ~~~ ({\rm for} ~ a\ell \sim fL_{\rm J}). \label{eq:asympto_rel}
\end{equation}
Here, the Jeans length is 
\begin{equation}
    L_{\rm J} = \sqrt{\frac{\pi c_{\rm s}^2}{G\rho}} \propto a^{3(2-\gamma_{\rm eff})/2}, \label{eq:Jeans}
\end{equation}
where $c_{\rm s}$ is the sound speed of a collapsing cloud.
The asymptotic value in the limit of $\rho \rightarrow \infty$ (or $a \rightarrow 0$)is
\begin{equation}
    \frac{\omega}{|H|} = \frac{5-3\gamma_{\rm eff}}{2\eta}. \label{eq:A_value}
\end{equation}

Next, we estimate $|H|(=\dot{a}/a=(1/3)(\dot{\rho}/\rho))$ in the collapsing gas core.
Since collapse proceeds in a self-similar fashion (see Appendix \ref{sec:similarity}), the growth of the central density is obtained by integrating a set of ordinary differential equations \citep{Yahil83,Suto88}. 
From the equations, the density $\rho(r,t)$ as a function of radius $r$ from the cloud center and time $t$ is given as $\rho(r,t)=\alpha(x)/(4\pi G t^2)$. $\alpha(x)$ is a dimensionless quantity as a function of dimensionless coordinate $x \equiv r/\sqrt{\kappa}t^n$, where $\kappa$ and $n$ are constants. See \citealt{Suto88} in detail.
At the center of the core $(x=0)$, the central density $\rho_{\rm c}$ is given as $\rho_{\rm c}=\alpha(0)/(4\pi G t^2)$, where $\alpha(0)$ is a constant depending on $\gamma_{\rm eff}$ (see Table \ref{tab:alpha}).
In a self-similarly collapsing gas core, we assume the gas density to be uniform ($\rho_{\rm mean}\simeq \rho_{\rm c}$), then we have 
\begin{equation}
    |H| = \left|\frac{1}{3}\frac{\dot{\rho}}{\rho}\right|=\left|\frac{2}{3}t^{-1}\right| = \frac{2}{3}\sqrt{4 \pi G \rho}\frac{1}{\sqrt{\alpha(0)}}. \label{eq:Hubble}
\end{equation}
Substituting this relation to Equation (\ref{eq:A_value}), we obtain
\begin{equation}
    \omega = \frac{5-3\gamma_{\rm eff}}{3\eta}\sqrt{\frac{4 \pi G \rho_{\rm mean}}{\alpha(0)}}. \label{eq:omega}
\end{equation}

Then, with $\omega = v/fL_{\rm J}$, we can derive the saturated turbulent velocity as,
\begin{equation}
    v_{\rm sat} = \frac{5-3\gamma_{\rm eff}}{3\eta}\frac{2\pi c_{\rm s}}{\sqrt{\alpha(0)}}f. \label{eq:vel_sat}
\end{equation}
Finally, we obtain the maximum saturation Mach number of turbulence 
\begin{equation}
    \mathcal{M}_{\rm sat} \simeq  \frac{5-3\gamma_{\rm eff}}{3\eta}\frac{2\pi}{\sqrt{\alpha(0)}}f. \label{eq:mach_sat}
\end{equation}
This expression depends on $\gamma_{\rm eff}$ but not on $\rho_{\rm mean}$.
This feature nicely agrees with the numerical results obtained in the previous section as we see in Section \ref{sec:comparison}.

\begin{table}[htbp]
 \caption{Calculated $\alpha(0)$ for each $\gamma_{\rm eff}$}
 \centering
   \begin{tabular}{l|cccccccc}
     \hline 
     $\gamma_{\rm eff}$ & 1.0 & 1.05 & 1.09 & 1.10 & 1.15 & 1.2 & 1.25 & 1.3 \\
     \hline 
     $\alpha(0)$ & 1.67 & 2.15 & 3.03 & 3.26 & 4.73 & 7.19 & 11.6 & 22.0\\
     \hline
   \end{tabular}
  \label{tab:alpha}
\end{table}
\section{Driving scale of turbulence}\label{sec:comparison}
In this section, we compare the numerical results in Section \ref{sec:results} with the analytic estimate in Section \ref{sec:estimate}.
First, we solve the ordinary differential equations (Equations 15a and b) of \cite{Suto88} by using the fourth-order Runge-Kutta method to obtain $\alpha(0)$, which is necessary to calculate $\mathcal{M}_{\rm sat}$.
We summarize obtained $\alpha(0)$ corresponding to each $\gamma_{\rm eff}$ in Table \ref{tab:alpha}.

Besides, we also need the dissipation coefficient $\eta$ of turbulence.
\cite{MacLow1999} performed the simulations of uniform, randomly driven turbulence and found that the driven turbulence saturates, resulting in the balance between energy injection and dissipation.
He also found that the energy injection rate and the dissipation rate of turbulence can be well approximated by the linear relation, and derived the value of the dissipation coefficient.
However, his simulations were performed only with the isothermal equation of state, so that the $\gamma_{\rm eff}$ dependence is still unclear.
Therefore, we perform another set of simulations, based on the simulations of \cite{MacLow1999}, to confirm the value and $\gamma_{\rm eff}$ dependence of $\eta$.
Consequently, we obtain $\eta = 0.42$ for any $\gamma_{\rm eff}$, which is the same as \cite{MacLow1999}, i.e. the dissipation coefficient has no $\gamma_{\rm eff}$ dependence.
Thus, we use this value to obtain the saturation Mach numbers from Equation (\ref{eq:mach_sat}).
The detailed setup and results of simulations are summarized in Appendix \ref{sec:turb}.

\begin{figure}[htbp]
 \centering
  \plotone{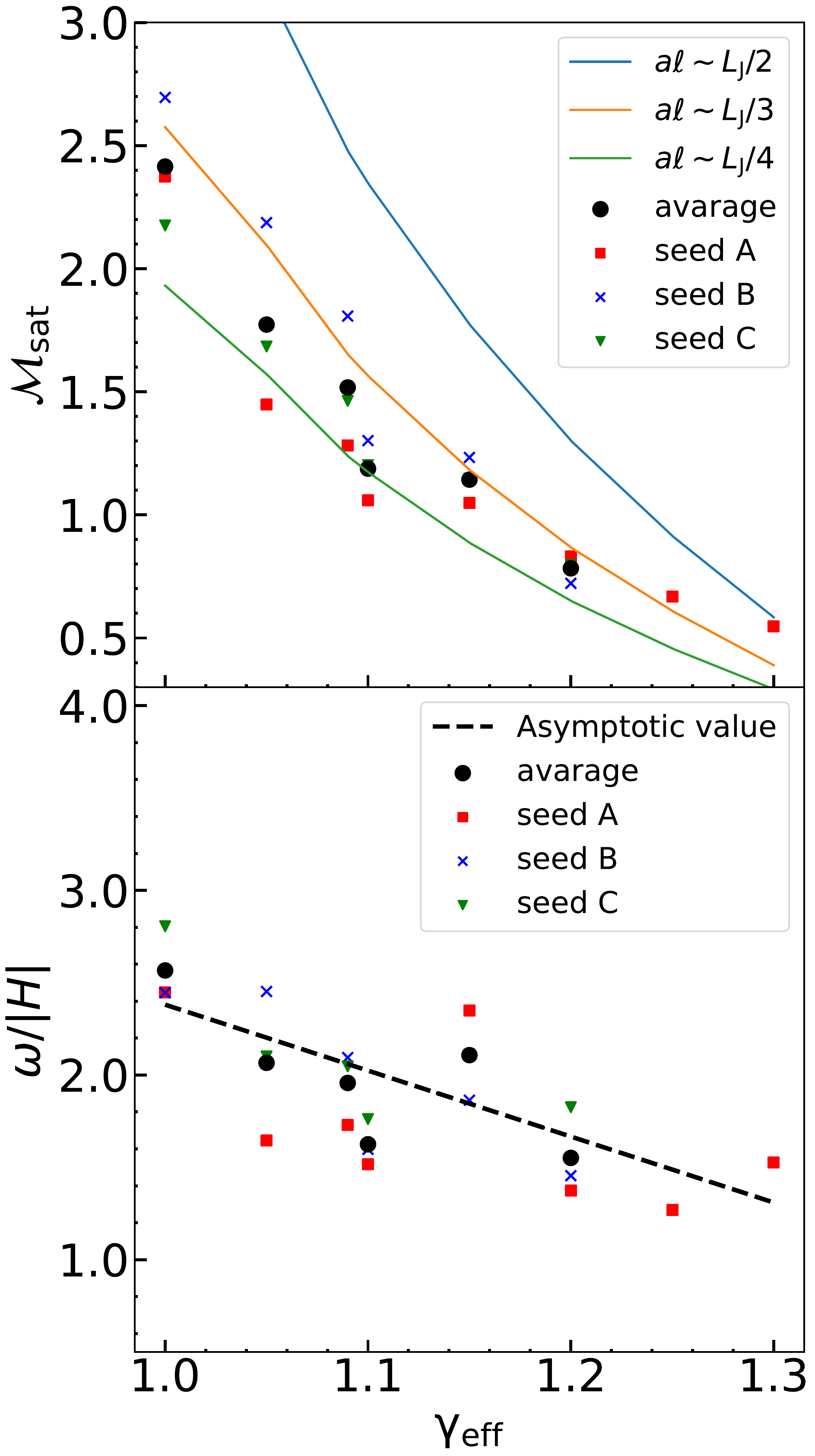}
  \caption{Saturation Mach number $\mathcal{M}_{\rm sat}$ v.s. $\gamma_{\rm eff}$ (upper panel) and ratio of eddy turnover frequency $\omega$ and the absolute value of Hubble parameter $|H|$ v.s. $\gamma_{\rm eff}$ (lower panel). The red square, blue cross, and inverse triangle symbols denote the numerical results for $\mathcal{M}_0 = 0.1$ with seeds \textbf{A}, \textbf{B}, and \textbf{C}, respectively. The blue, orange, and green solid lines in the upper panel are estimated saturation Mach numbers using Equation (\ref{eq:mach_sat}) which assumes turbulence driving scale $a\ell \sim L_{\rm J}/2$, $L_{\rm J}/3$, and $L_{\rm J}/4$, respectively.
  The black dashed line in the lower panel is obtained from Equation (\ref{eq:A_value}).
  }
    \label{fig:saturation} 
\end{figure}

The upper panel of Figure \ref{fig:saturation} shows the saturation Mach numbers for each $\gamma_{\rm eff}$ calculated from Equation (\ref{eq:mach_sat}).
The red square, blue cross, and green inverse triangle symbols are obtained from our numerical results for $\mathcal{M}_0 = 0.1$ with the initial seeds \textbf{A}, \textbf{B}, and \textbf{C}, respectively.
The black circle symbols are the average of the three models. 
As shown in Section \ref{sec:results}, the turbulent velocities saturate at $\rho_{\rm mean} \gtrsim 10^{-13} ~ \mathrm{g ~ cm^{-3}}$ in all models.
Therefore, we regard the time-averaged Mach number in $\rho_{\rm mean} \geq 10^{-11} ~ \mathrm{g ~ cm^{-3}}$ as the saturation Mach number $\mathcal{M}_{\rm sat}$.
The blue, orange, and green solid curves denote $\mathcal{M}_{\rm sat}$ estimated from Equation (\ref{eq:mach_sat}) with turbulence driving scales of $a\ell \sim L_{\rm J}/2$, $L_{\rm J}/3$, and $L_{\rm J}/4$, respectively.

We can see that, if the turbulent driving scale is one-third of Jeans length, the average of the three initial seeds (black circle symbols) are in good agreement with the theoretical estimate of the saturation Mach number.
This is also almost consistent with the result of \cite{Guerrero-Gamboa20} for an isothermal core, who obtained the driving scale of 0.285 $L_{\rm J}$ with the same dissipation coefficient as \citet{MacLow1999} but with a different theoretical model based on virial equilibrium of a core.

The lower panel of Figure \ref{fig:saturation} shows the ratio of eddy turnover frequency $\omega$ and the absolute value of Hubble parameter $|H|$ as a function of $\gamma_{\rm eff}$ in the saturation regime.
The symbols indicate the results obtained with the same method as the upper panel, and the black dashed line is obtained from Equation (\ref{eq:A_value}).
Unlike the solid curves in the upper panel, the dashed line in the lower panel has a weaker dependency against $\gamma_{\rm eff}$ because Equation (\ref{eq:A_value}) does not contain $\alpha(0)$.

As well as the saturation Mach number, we can see that the average of the results with the three initial seeds (black circle symbols) are in good agreement with the theoretical estimate.
These results provide clear evidence that the saturation mechanism of turbulence is a balance between the amplification and the decay of turbulence.

Important point of the present study is that our theory can well describe the numerical results of various $\gamma_{\rm eff}$.
This versatility of the theory supports its significance.

\section{Discussion}\label{sec:discussion}

\subsection{On the PopIII formation}\label{sec:popIII}

In this paper, we have focused on the turbulence in the runaway collapse phase.
As a result, given an effective polytropic index $\gamma_{\rm eff}$,
we find an analytic formula to assess the saturation level of turbulence in collapsing cores.
In case we consider the primordial star formation which is well approximated by $\gamma_{\rm eff}\simeq 1.09$,
the turbulence will be supersonic $\mathcal{M}_{\rm sat}\sim 1.5-2$ (Figure \ref{fig:saturation} and Equation \ref{eq:mach_sat}),
which is consistent with the results of earlier works \citep[e.g.,][]{Greif2012}.
Therefore, supersonic turbulence should be a general feature of the collapsing star-forming clouds in primordial environments.

Not only turbulence but also magnetic fields can have substantial impacts on the IMF of the first stars if they are sufficiently strong.
It has been considered that magnetic fields are very weak in the early universe, but
recent high-resolution simulations have revealed that
they are amplified to certain levels by the small-scale dynamo effect due to turbulence \citep[e.g.,][]{Sur10,Sur12,Fed11,Turk2012,Sharda21_dynamo}.
Thus, the magnetic energy can be potentially amplified to the level of the turbulent energy.
We can estimate the maximum magnetic field strength as $B_{\rm max}\sim 1.6 \times 10^4 ~ {\rm G}$ for $n_{\rm H} = 10^{19} ~ {\rm cm}^{-3}$, assuming that the magnetic energy can reach the turbulent energy estimated from the results in Section \ref{sec:results}, which in turn will affect the dynamics of the infalling gas.

The accretion phase, which is not investigated in this paper, is crucial for the Pop III formation.
Recent numerical studies have indicated that both the turbulence and magnetic field play important roles in the accretion phase, for instance, a fragmentation of the accretion disk is promoted/suppressed by their effects, whereas the impact on the fragmentation depends on their strength  \citep[e.g.,][]{Wollenberg20,Sharda20,Sharda21_dynamo,Prole22}.
In this paper and \citetalias{HSC21}, we have found that the amplification of turbulence always occurs in the collapsing primordial gas clouds and the supersonic turbulence is developed.
Given that the small-scale dynamo is extremely efficient at generating strong magnetic fields from turbulence, both strong turbulence and magnetic fields likely always exist at the onset of the accretion phase and should be included in the simulations of accretion phase for the first star formation.

\subsection{On the present-day star formation}\label{sec:present-day}
In the primordial case, the relation between the gas density and pressure is described by a single power law index $\gamma_{\rm eff}$.
On the other hand, the temperature evolution in the present-day star formation is more complicated.
The temperature evolution from cloud collapse to star formation 
tracks 4 phases: 
(i) isothermal collapse ($\gamma_{\rm eff}=1.0$, $\rho \lesssim 10^{-14}~\mathrm{g ~ cm^{-3}}$),
(ii) first-core formation ($\gamma_{\rm eff}=1.4$, $10^{-14}~\mathrm{g ~ cm^{-3}}\lesssim \rho \lesssim 10^{-8}~\mathrm{g ~ cm^{-3}} $),
(iii) the second collapse ($\gamma_{\rm eff}=1.1$ , $10^{-8}~\mathrm{g ~ cm^{-3}}\lesssim \rho \lesssim 10^{-4}~\mathrm{g ~ cm^{-3}} $),
(iv) protostar formation ($\gamma_{\rm eff}=1.67$, $10^{-4}~\mathrm{g ~ cm^{-3}}\lesssim \rho$) \citep[e.g.,][]{Masunaga00a,Tomida13}.
The gas is almost isothermal until the gas density reaches $10^{-14} ~\mathrm{g ~ cm^{-3}}$,
and the turbulent velocity is already transonic in the molecular clouds at much lower densities.
Therefore, the saturation level of turbulence can be estimated as $\mathcal{M}_{\rm sat}\sim 2-3$ just before the first core formation according to Fig \ref{fig:saturation} and Equation (12).
From the first core phase, the equation of state becomes much stiffer,
which may result in lower $\mathcal{M}_{\rm sat}$.
It is difficult to predict the evolution of turbulence when $\gamma_{\rm eff}$ changes during the collapse,
because we have investigated just for the single power index $\gamma_{\rm eff}$.
We will investigate the effects of such a complex equation of state in future works.

From the observational side, 
high-resolution instruments, such as ALMA, have begun to provide insights on the effects of turbulence on star formation, targeting
dense cores in molecular clouds.
For example, several authors reported multiple systems with large separation and highly misaligned rotational axes, indicating the fragmentation due to turbulence during the protostar formation \citep{Williams14,Fernandez-Lopez17,Lee17}.
Besides, observations of Taurus by \cite{Tokuda18} indicate that intense turbulent shock heating possibly occurs in star-forming cores.
Our results may indicate that strong turbulence at such small scales is driven by gravitational collapse and in turn, it causes the fragmentation and the shock waves.
It will be clarified whether such turbulence is actually driven by gravitational collapse or not by further high-resolution observations of gravitationally unstable ``starless'' dense cores in the future.

\subsection{A caveat on the initial condition}\label{sec:limit}
In this study and \citetalias{HSC21}, we initially give the velocity fluctuation which has a characteristic energy spectrum ($E(k) \propto k^{-2}$).
Strictly speaking, this initial condition mimics only the velocity field of turbulence in realistic gas clouds, but not the density field that should be correlated with the velocity.
In order to develop this ``real'' turbulence, twice the large eddy turnover times are necessary at least. In contrast, the eddy turnover timescale is initially longer than the collapse timescale ($v_{\rm turb} < c_{\rm s}$) as shown in Section 4.6 in \citetalias{HSC21}. As a result, the cloud collapse proceeds before the initial field well develops into a turbulent flow.  Thus we have to keep in mind this incompleteness of the initial field.
However, considering the cosmological simulations that naturally generate the initial gas distribution also obtain similar growth/saturation of the velocity field \citep[e.g.,][]{Greif2012}, we guess that the numerical results from the ``real'' initial turbulence will be similar to the results in the present paper.

\section{Summary}\label{sec:summary}
In this paper, we have studied the saturation mechanism of amplified turbulence in collapsing gas clouds.

First of all, assuming the polytropic equation of state, we perform high-resolution collapse simulations with various polytropic exponents $\gamma_{\rm eff}$ to investigate the evolution of the turbulent velocity.
We confirm that turbulence can be amplified through gravitational collapse and its saturation level depends on $\gamma_{\rm eff}$, as found in \citetalias{HSC21}.
We also perform collapse simulations with various initial velocities, such as high Mach number and turbulence with initial cloud rotation, and find that there is no significant difference in the saturation level.
To investigate the effects of baroclinic term leading to the time variation of turbulent vorticity, we follow cloud collapse until protostar formation, self-consistently solving chemical reactions in primordial gas to introduce radiative cooling explicitly.
We find that the baroclinic term due to shock waves/radiative cooling has a quite small effect on the evolution of turbulence.
Finally, we analytically estimate the saturation level of turbulence by using similarity solutions and the turbulence dissipation coefficient $\eta$.

We find that the numerical results are well described by the analytic formula with the turbulence driving scale of one-third the Jeans scale.

We emphasize that the present results show the saturation level of turbulence in contracting prestellar cores is given by the balance between the 
energy input rate from gravitational collapse
and the dissipation rate of turbulence. 
The former depends only on $\gamma_{\rm eff}$ (Equation \ref{eq:asympto_rel}) while the latter is a constant (see Appendix \ref{sec:turb}).
Therefore, the strength of turbulence at the epoch of first-core formation can be estimated solely by $\gamma_{\rm eff}$ both in the early and present-day universe.

\software{\texttt{Enzo} \citep{Enzo,Brummel-Smith19}, \texttt{Grackle} \citep{grackle}, \texttt{yt} \citep{yt}}

\acknowledgments
We are grateful to anonymous referee for careful reading of the manuscript and constructive comments.
We thank T. Inoue, K. Tomida, K. Omukai, T. Hosokawa, M. Machida, K. Sugimura, and M. I. N. Kobayashi for fruitful discussions and useful comments. We are thankful for the support by Ministry of Education, Science, Sports and Culture, Grants-in-Aid for Scientific Research No. 22K03689, 17H02869, and 17H06360.
This work was supported by JST SPRING, Grant Number JPMJSP2117.
A part of numerical calculations in this work was carried out on Yukawa-21 at the Yukawa Institute Computer Facility.
Computations and analysis described in this work were performed using the publicly-available \texttt{Enzo} , \texttt{Grackle}, and \texttt{yt} codes, which is the product of a collaborative effort of many independent scientists from numerous institutions around the world.

\appendix
\section{Comparison with similarity solution}\label{sec:similarity}
Assuming that the evolution of the density in a gas cloud follows the similarity solution 
, we have analytically estimated the saturation Mach number in Section \ref{sec:estimate}.
Although we have shown that this assumption is reasonable in three-dimensional collapse simulations for $\gamma_{\rm eff}=1.09$ \citepalias{HSC21}, we did not confirm its validity for other $\gamma_{\rm eff}$.
Therefore, we compare our simulation results with the similarity solution for all the $\gamma_{\rm eff}$ that we consider.

\begin{figure}[htbp]
 \centering
  \plotone{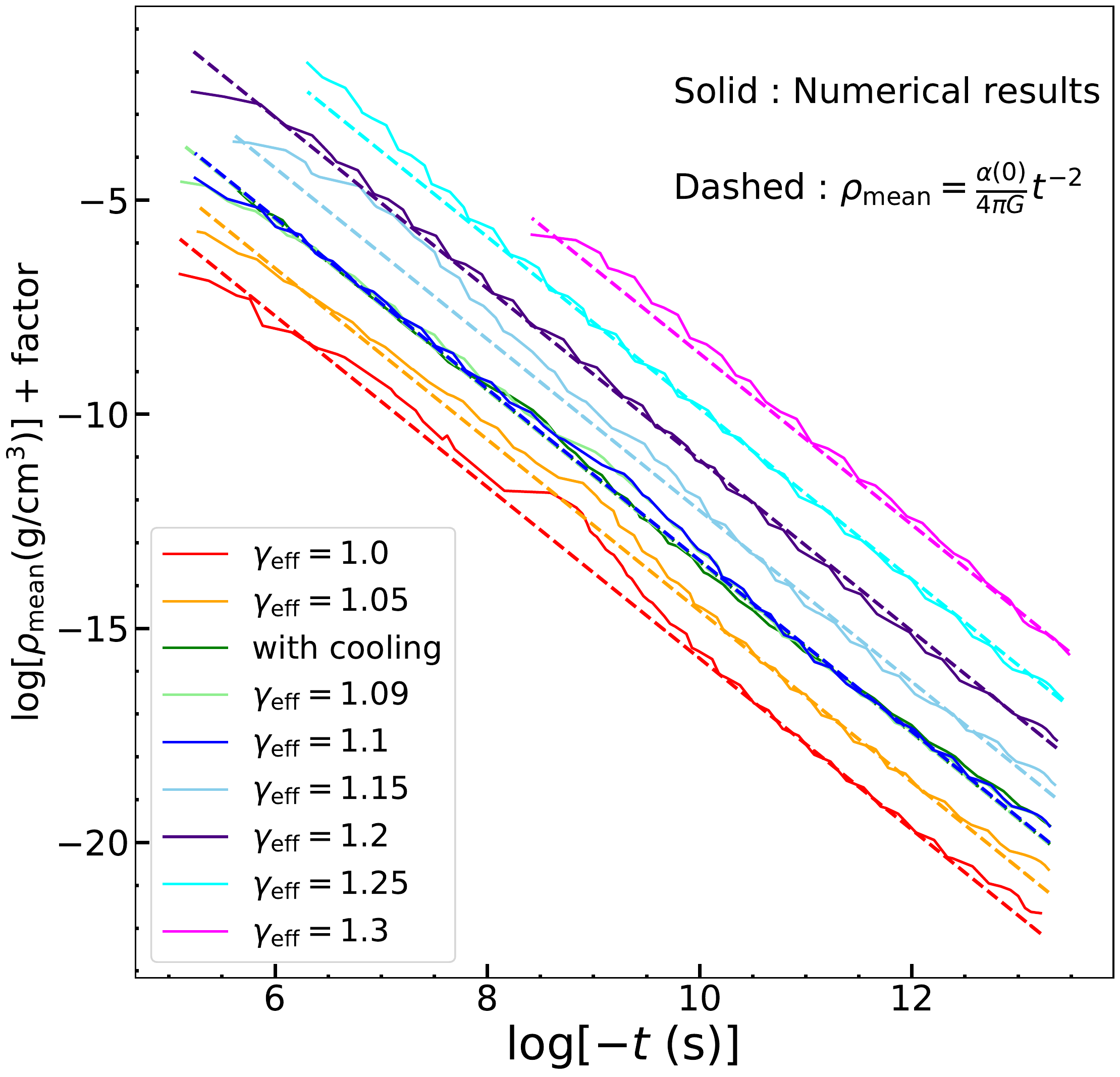}
    \caption{Evolution of the core mean densities as a function of time $-t$. The solid lines are numerical results. The dashed lines which have corresponding colors denote $\rho_{\rm mean}=(\alpha(0)/4\pi G) t^{-2}$, where $\alpha(0)$ are given in Table \ref{tab:alpha}. Note: These values except for `with cooling', $\gamma_{\rm eff}=1.09$, and $\gamma_{\rm eff}=1.1$ are multiplied by factors for ease of viewing.}
    \label{fig:time_dens} 
\end{figure}

Figure \ref{fig:time_dens} shows that the evolution of the mean core density as a function of negative time, where the origin $t=0$ is the time of the final snapshot.
All the curves except for `with cooling', $\gamma_{\rm eff}=1.09$, and $\gamma_{\rm eff}=1.1$ are multiplied by factors for ease of viewing.
Although the density evolution becomes faster due to the cloud deformation by the turbulence for $\gamma_{\rm eff}=1.0$ and 1.15, it can be clearly seen that the density evolution in the three-dimensional simulations (solid curves) and the similarity solution (dashed curves) is almost in good agreement for all $\gamma_{\rm eff}$.
This result supports that our assumption in Section \ref{sec:estimate} is reasonable.

\section{Simulations of turbulence dissipation}\label{sec:turb}
To estimate $\mathcal{M}_{\rm sat}$ from Equation (\ref{eq:mach_sat}), we need the dissipation coefficient $\eta$ of turbulence.
%
In this appendix, we extend the simulations of \cite{MacLow1999}, who investigated turbulent dissipation for isothermal gas.
We perform simulations for various $\gamma_{\rm eff}$, based on the method of \cite{MacLow1999}.
Here we remark that 
the dissipation coefficient $\eta_{\rm v}$ defined by \cite{MacLow1999} is slightly different from $\eta$ in Equation (\ref{eq:mach_sat}).
They can be converted to each other with the relationship of $\eta=2\pi \eta_{\rm v}$ \citep[equation 14 of][]{Guerrero-Gamboa20}.

\begin{figure}[htbp]
 \centering
 \plotone{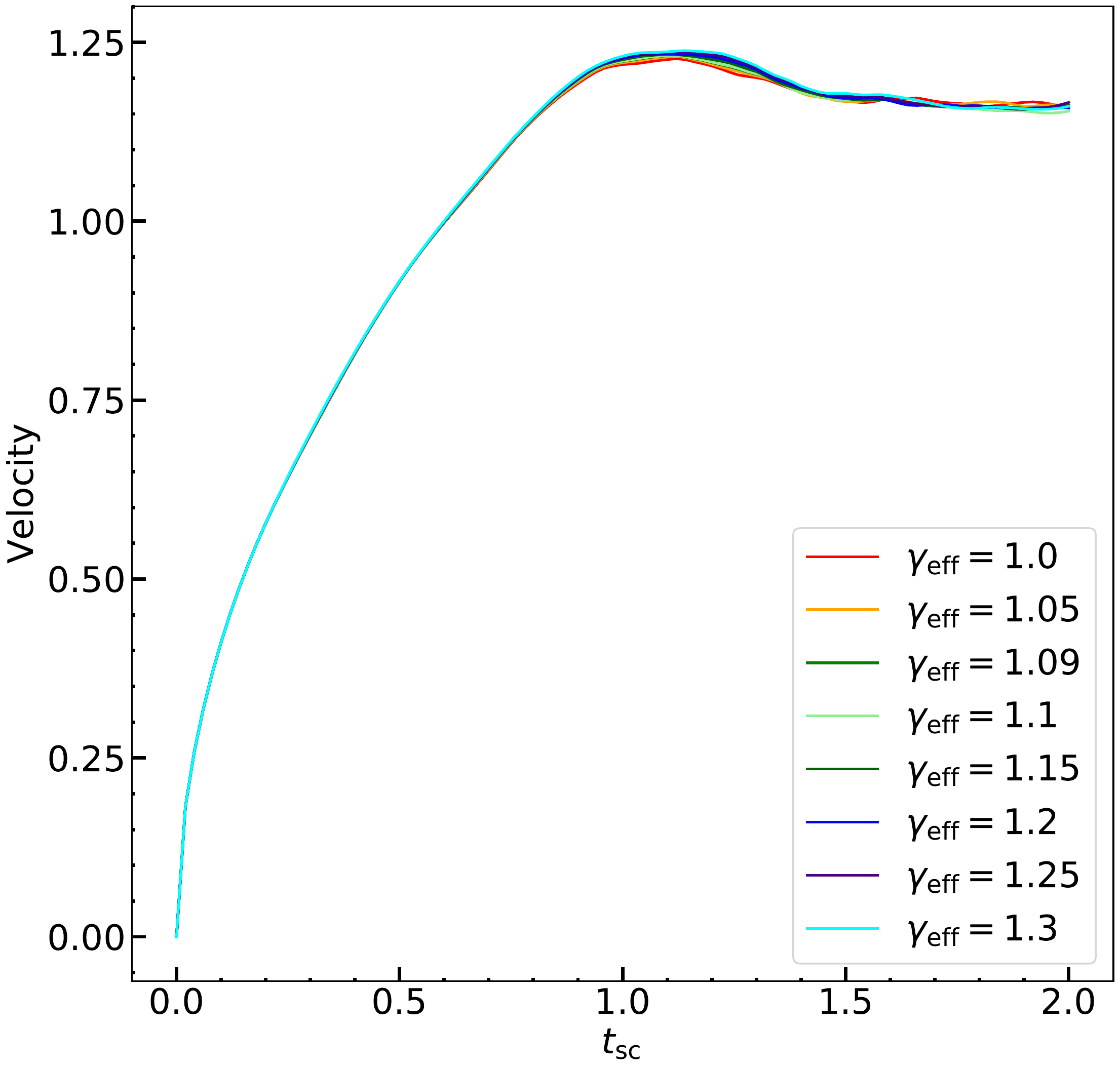}
  \caption{Temporal evolution of the turbulent velocity in the case with a kinetic energy input rate $\dot{E}_{\rm in} = 1$. The color of the curves corresponds to each $\gamma_{\rm eff}$ indicated by the box in bottom right.
  We can see no $\gamma_{\rm eff}$ dependence.}
    \label{fig:t-v} 
\end{figure}

First, we note that each physical quantity in this simulation is described using computational code units. 
All 
runs are performed on a three-dimensional uniform grid with $128^3$ 
cells.\footnote{\cite{MacLow1999} showed that, if enough resolution ($128^3$ at most) is provided, the growth of velocity in this kind of turbulence simulations does not depend on resolution. Hence we choose $128^3$ uniform grids.
Also, the resolution is comparable to the effective resolution of the collapse simulations in Section \ref{sec:collapse}.}
We set the computational domain length $L=1$ with a periodic boundary condition.
The gas is uniformly distributed in the domain with the initial density $\rho_0=1$ and the total mass $m\equiv \rho L^3=1$.
For comparison with the results of Section \ref{sec:baro}, we set $\mu=1.22$ and $\gamma_{\rm ad}=1.4$ and perform simulations with the same set of $\gamma_{\rm eff}$.

During the simulations, turbulence is continuously driven by the injection of velocity fluctuations $\delta v$ at each time step $\delta t$.
This $\delta v$ corresponds to a constant kinetic energy input rate $\dot{E}_{\rm in}=\Delta E / \Delta t$, which is set as an initial parameter.
The turbulent kinetic energy spectrum at a wavenumber $k$ is set to follow the Larson's law (i.e., $E(k)\propto k^{-2}$), which has a peak at $k_{\rm peak}=2$.
Thus, the typical driving scale of turbulence is $L/k_{\rm peak}$.
The driven velocity field is composed of fully solenoidal mode, which mimics the turbulent velocity field driven by gravitational collapse \citep[][]{Fed11,HSC21}.
We perform simulations with $\dot{E}_{\rm in}= 0.1, ~ 0.3, ~ 1, ~ 3, ~ 10$ for every $\gamma_{\rm eff}$ and terminate them at twice the sound crossing time $t_{\rm sc}$ calculated by the initial sound speed given by each $\gamma_{\rm eff}$.
\begin{figure}[htbp]
 \centering
 \plotone{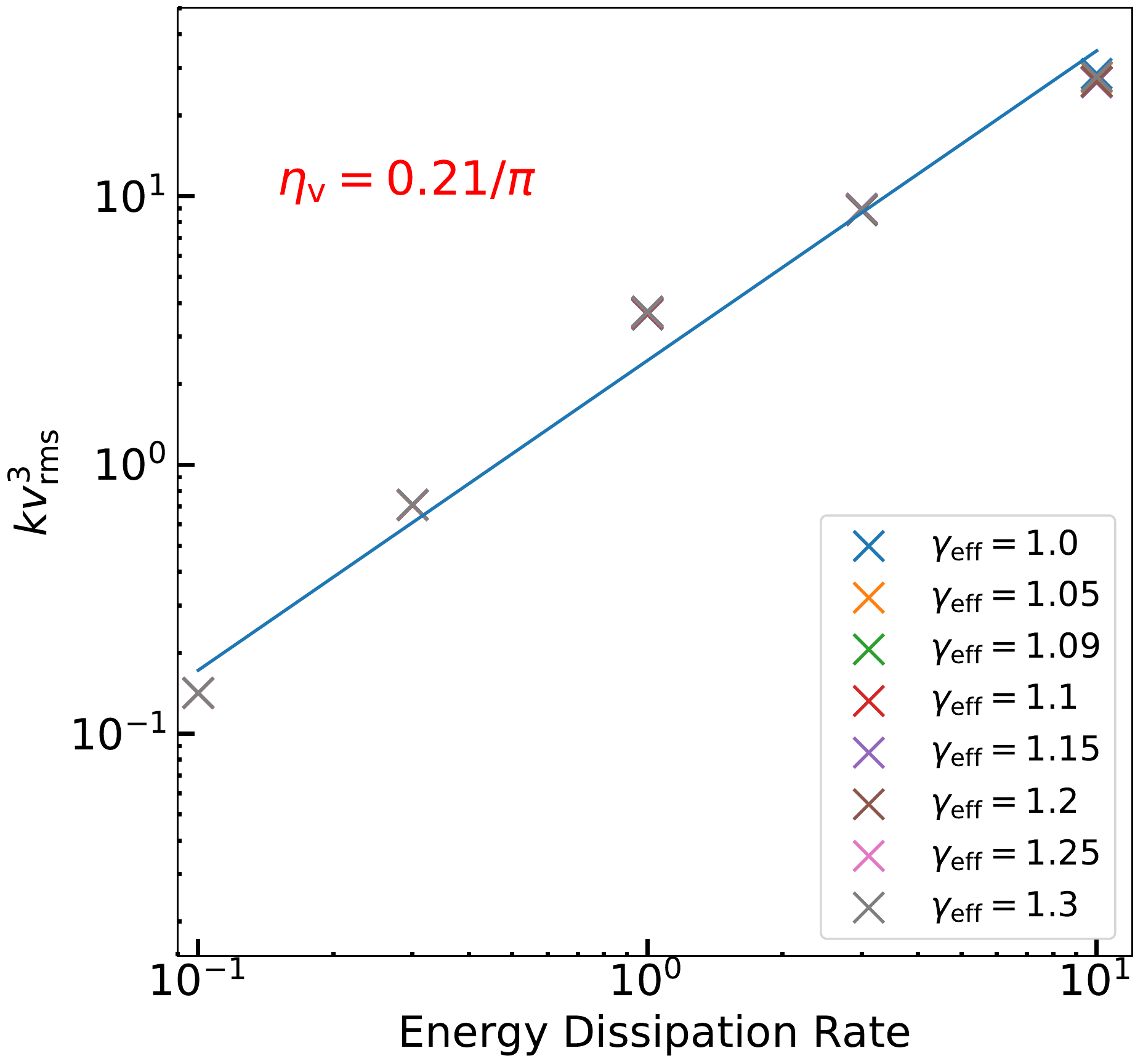}
  \caption{Scatter plot for the energy dissipation rate ($=\dot{E}_{\rm in}$) v.s. $kv_{\rm rms}^3$. Difference of colors corresponds to the difference of $\gamma_{\rm eff}$. The blue solid line is obtained by a least-squares method. Note: There is no difference in values among different $\gamma_{\rm eff}$, so that, the points are almost degenerated to a single point at each energy dissipation rate.}
    \label{fig:Ediss} 
\end{figure}

First, as an example, we show the temporal evolution of the turbulent velocity for $\dot{E}_{\rm in}=1$ in Figure \ref{fig:t-v}.
The velocity increases quickly and saturates at $\simeq t_{\rm sc}$ for all $\gamma_{\rm eff}$.
We can see that the evolution of velocity is almost the same for every $\gamma_{\rm eff}$, which indicates that there is no $\gamma_{\rm eff}$ dependence during the growth of turbulence. 

\cite{MacLow1999} showed that, when the turbulence reaches an equilibrium state, the energy dissipation rate $\dot{E}_{\rm kin}$ of turbulence equals to the energy injection rate $\dot{E}_{\rm in}$ and 
can be well approximated by the linear relation
\begin{equation}
    \dot{E}_{\rm kin} \simeq -\eta_{\rm v}m \Tilde{k} v_{\rm rms}^3, \label{eq:diss_rate}
\end{equation}
where $\Tilde{k}\equiv (2\pi /L)k_{\rm peak}$ 
is a ``dimensionalized wavenumber''. 
Using this relation, we estimate the dissipation coefficient.

We plot $kv_{\rm rms}^3$ v.s. energy dissipation rate ($=\dot{E}_{\rm in}$) after the sound crossing time from the beginning of the simulation in Figure \ref{fig:Ediss}.
The circles in different colors denote the results for corresponding $\gamma_{\rm eff}$.
We can see that there is no difference among different $\gamma_{\rm eff}$, and
the points are almost overlapped at a single point for each energy dissipation rate.
From the results and Equation (\ref{eq:diss_rate}), we can estimate $\eta_{\rm v} = 0.42/(2\pi)$ with the least squares method.
It is fully consistent with the result of the isothermal case in \cite{MacLow1999}.
Hence we use $\eta=0.42$ in section \ref{sec:comparison}.

\bibliography{references}{}
\bibliographystyle{aasjournal}
\vspace{5mm}

\listofchanges
\end{document}